\documentstyle[12pt,epsfig]{article}

\def\Title#1{\begin{center} {\Large {\bf #1} } \end{center}}

\begin{document}

\Title{Gravity, Particle Physics and their Unification\footnote{
Talk given at Lepton-Photon 99.}}

\bigskip\bigskip


\begin{raggedright}  

{\it J. M. Maldacena\index{Maldacena, J. M.}\\
Department of Physics \\
Harvard University, 
Cambridge, Massachusetts 02138 }
\bigskip\bigskip
\end{raggedright}

\section{Introduction}

Our present world picture is based on  two theories: the
standard model of particle physics and general relativity, the theory
of gravity. These two theories have scored astonishing successes.
It is therefore quite striking when one learns 
that this picture of the laws of physics
is inconsistent. The inconsistency comes from taking a part of theory,
the standard model, as a quantum theory while the other, gravity, 
as a classical theory. At first sight it seems that all we need to do
is to 
quantize general relativity. If one performs an expansion in terms of
Feynman diagrams, the technique used in the standard model, then one
finds infinities that cannot be absorbed in a renormalization of the 
Newton constant (and the cosmological constant). In fact as one goes
to higher and higher orders in perturbation theory one would have
to include more and more counterterms. In other words, the theory is
not renormalizable. So the principle that is so crucial for 
constructing the Standard model fails. 
It is now more surprising that an inconsistent theory could agree
so well with experiment!. What happens is that quantum gravity effects
are usually very small due to the weakness of gravity relative to 
other forces. Since the effects of gravity are proportional to the mass,
or the energy of the particle, they grow at high energies. At energies
of the order of $ E \sim 10^{19} Gev$ gravity would have a strength 
comparable with that of the other Standard Model interactions\footnote{
This is an energy scale where we should definitely see new physics, it is 
nevertheless possible  that quantum gravity becomes relevant
at much lower energies, $E \sim $ 1-10 $Tev$ \cite{Arkani-Hamed:1998rs}. }. 
We should also remember that physics as we understand it now cannot
explain the most important ``experiment'' that ever happened: the Big 
Bang.  It is quite interesting that the Big-Bang theory links
high energy physics and cosmology and in order to understand what 
happened in the beginning it seems that we need 
 to understand quantum gravity. 
There are also esthetic reasons for wanting a theory beyond the 
Standard Model. We would like to explain the origin of the gauge 
group, the relations between the three couplings, the rest of the 
parameters of the standard model, why we have three generations, etc. 
It is very suggestive
that if one extrapolates the running of the couplings they seem to 
meet at energies that are close to the energy where quantum gravity
becomes important, suggesting that a grand unified field theory, based
on a bigger gauge group would in any case lie close to the 
Planck scale. 
The present situation is analogous to the situation particle physics
was in when we only had Fermi's theory of weak interactions. It was a
theory that agreed well with experiments performed at low energies but
it was not a consistent theory. Renomalizability, or mathematical 
consistency, was a crucial clue for the discovery of the Standard 
Model. 

The challenges we face  can be 
separated, in degree of difficulty, in the following three: 

I) Formulate an internally consistent theory of quantum gravity. 
By this we mean a theory which reduces at low energies, $ E \ll 10^{19}
Gev$,
to general relativity but in which we can perform quantum calculations
to any order we wish. This theory should solve some fundamental 
problems with quantum gravity such as explaining the origin of black 
hole entropy, etc. These are questions about gravity which do not
involve directly the fact that we also have the particle physics that
we see in nature.

II) Be capable of incorporating the Standard Model. So the theory should
be such that at low energies it can   contain chiral gauge fields,
fermions, etc. 

III) Explain the Big-Bang and the parameters of the Standard Model. 
We should understand the resolution of the initial singularity in 
cosmology and we should understand why we have the Standard Model. 
We should understand how the standard model parameters arise, 
which parameters are related, and which parameters (if any) arise as
a ``historical'' accident. 

We know have a theory, called string theory (or M-theory), which
 has been able already to provide a solution to the first
two challenges. Unfortunately we do not know yet how to solve the
third challenge. Maybe  string theory is the solution and we just 
have to understand it better or maybe  we have to modify it in some
way. String theory is a theory under construction. We know several limits
and aspects of the theory but we still do not know the fundamental 
axioms of the theory that would enable us to approach the third challenge. 

\begin{figure}[htb]
\begin{center}
\epsfig{file=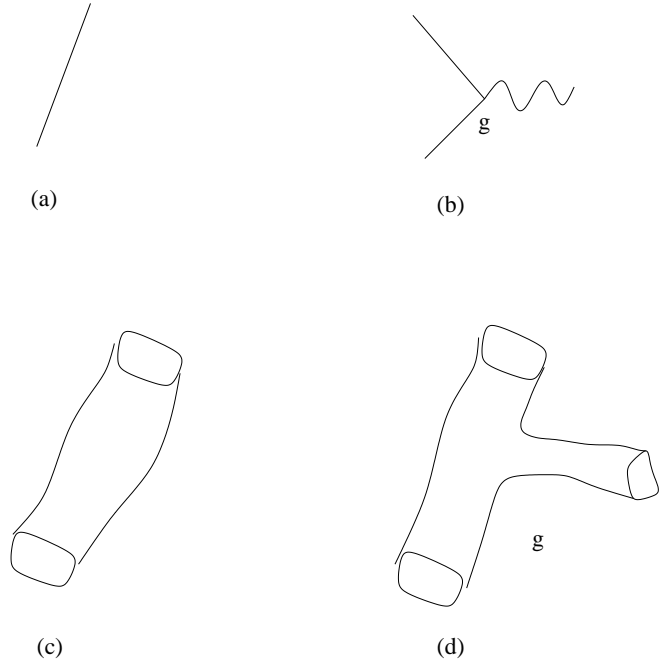}
\caption{Feynman diagrams for strings and particles. (a) Free particle 
propagation. (b) Interaction vertex between particles. (c) Free string
propagating. (d) String interacting with other strings.}
\label{fig:diagrams}
\end{center}
\end{figure}

\begin{figure}[htb]
\begin{center}
\epsfig{file=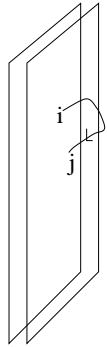}
\caption{D-branes are extended solitons whose excitations are described
by open strings which end on the brane. A string going from brane $i$ to 
brane $j$ is labeled by two indices. These give rise to non abelian 
gauge fields living on the D-brane. }
\label{fig:dbrane}
\end{center}
\end{figure}

String theory is based on the idea that fundamental objects are not
point particles, as in particle theories, but one dimensional objects
called
strings. 
Let us first review how we construct a theory of interacting particles.
We  start with a set of free particles, for example electrons,
photons,
quarks, gluons. These particles can have different states; e.g. they could
have the spin pointing up or down. 
 Then we consider  interactions. These are 
introduced by allowing particles to split into two other particles
with some probability amplitude $g$. $g$ will be the interaction strength
or coupling constant. For example and electron could emit a photon, etc. 
So in order to compute a scattering amplitude we have to sum over 
all paths of the particles and all ways in which they could emit other
particles, etc. These sums are performed via Feynman diagrams. 
In figure \ref{fig:diagrams}(a,b) we see some examples of Feynman diagrams. 
String theory is constructed in a completely analogous way. We first start
with free strings. Strings can be open or closed. Let us just consider
a theory with only closed strings. The strings are ``relativistic'',
meaning that their tension is equal to their mass per unit length. So
that if we had a stretched string an oscillation would propagate at 
the speed of light along it. The tension is a dimensionful quantity which
we can parameterize in terms of a distance scale $T = 1/l_s^2$. 
Strings can oscillate. These oscillations can
be decomposed in normal modes. Since strings are quantum mechanical 
objects each normal mode will have a certain occupation number. The 
total energy of the oscillating string will be quantized. And the 
total mass of an oscillating string will be equal to the total energy
contained in the oscillations. When we view this oscillating string from
far away it looks like a pointlike object. These different oscillatory
states of the string are analogous to the different polarization 
states of the particles, now  that the mass of the string state  itself 
depends on 
the  ``polarization'' state.  
 Some of these oscillatory states of the string
will have zero energy and will thus be  massless particles.
There is one state with spin two which can  be viewed as the  graviton. 
The masses of the massive string states are of the order $m \geq 1/l_s$.  
String interactions are introduced by allowing strings that touch to 
recombine into one string. These are splitting and joining interactions
as shown in figure \ref{fig:diagrams} (c,d). The amplitude for this process
defines the string coupling $g$. In order to compute any process in 
string theory we have to sum over all possible splitting and joining
interactions. 
The simplest string theories are those that live in ten dimensions and 
are supersymmetric. The sum over string theory Feynman diagrams can be
performed and yields finite results. At low energies, energies 
lower than the mass of the massive string states $E \ll 1/l_s$ the 
only excitation we will have are gravitons and other massless particles.
The interactions of these particles are those of Einstein gravity 
plus some other massless fields.
In this way string theory manages to quantize gravity. 
What we have described here amounts to a perturbative quantization of
the theory, in the same way that the Feynman diagram expansion in 
particle physics is a 
perturbative quantization of the field theory. 
But there are non-perturbative aspects of the theory that are not
captured by the perturbative theory. 
One example is soliton solutions, like magnetic monopoles of grand
unified theories. These are collective excitations that are stable,
typically for some topological reason. Their masses go as $m \sim 1/g^2$
and
in the weak coupling approximation
we can study them as solutions of the classical field theory action.
In field theories we can also have extended solitons, like cosmic strings
or domain walls.  
In string theory we also have solitons. These solitons are called
D-branes. D-branes are solitons with different dimensionalities. They
can be pointlike (D-0-brane), one dimensional (D-1-brane), two-dimensional
(D-2-brane), etc.
These solitons have a very precise description in string theory
\cite{Polchinski:1995mt}.
Their excitations are described by open strings that end on them. 
At low energies some of the open string modes are massless, have 
spin one and give rise to gauge fields. When we put many branes
together the open strings have two indices $i,j$ labeling the brane where
they start and the brane where they end see figure \ref{fig:dbrane}.
 These two indices become 
the indices of non-abelian $U(N)$ gauge fields. 

\begin{figure}[htb]
\begin{center}
\epsfig{file=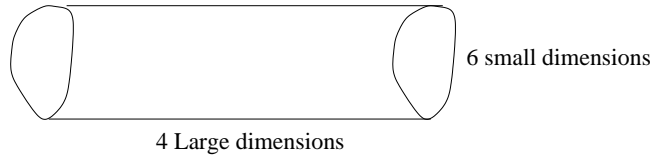}
\caption{Picture of a ten dimensional geometry of the form $R^4 \times M^6$
where $M^6$ is a small six dimensional space and $R^4$ represents the four
large dimensions that we see, 3 spatial and time. }
\label{fig:compactification}
\end{center}
\end{figure}

It might be surprising that we were discussing a ten dimensional theory
while our world is ``obviously'' four dimensional. 
What we really see is that the world is four dimensional at long distances,
we do not really 
know the dimensionality of the world at short enough distances. 
In string theory we assume that we are living in a world that has four
large dimensions (the ones we see) and six very small dimensions, see
figure \ref{fig:compactification}.
It is a familiar phenomenon in condensed matter that if an electron is
confined to move on a very narrow layer then the electron will behave
as if it was  moving in only two dimensions. Similarly particles  that 
move in a ten dimensional space with six small dimensions will behave 
at low energies as if they were moving only in four dimensions. 
What is the size of these extra dimensions?  The traditional
view is that they are  all small, of the order of $10^{-33} cm$.
 But recently it was realized  that 
some dimensions could be  as big as 
$1 mm$  \cite{Arkani-Hamed:1998rs}. 
In that case all the standard model fields would have to 
be confined to live on a D-brane that is extended along the four 
extended dimensions that we see but transverse to the large extra dimensions. 
 Picking different
manifolds or brane configurations 
we can have different particles at low energies.
 In both cases the Standard Model parameters would depend on the details
of the internal manifold or brane configuration.
Compactifications which preserve 8,4 or 2 supersymmetries at low energies
are fairly well understood. The case where we preserve only one 
supersymmetry is not so well understood and we do not understand how 
supersymmetry can be broken, as it is in the real world, without generating
a huge cosmological constant, of the order of the supersymmetry breaking
scale. This seems to be the most important obstacle in understanding 
precisely how the Standard Model is embedded in string theory. 

Recent progress in string theory was based on the idea of 
dualities. It is familiar that classical electromagnetism
is invariant under the interchange of electric and magnetic fields
$\vec E \to  \vec B$, $\vec B \to - \vec E$. This exchanges electric 
charges with magnetic charges. In field theories electric charges are
carried by fundamental particles and magnetic charges by solitons. 
So this duality exchanges elementary particles with solitons. 
This can be achieved by changing the coupling constant as $g \to 1/g$ so 
that solitons that were heavy become light, like elementary particles. 
In many string theories we have dualities of this type. When
the coupling becomes strong in terms of some variables the theory
has an equivalent description in terms of some dual variables that
can be weakly coupled. In this fashion many string theories are connected.
These dualities are hard to check since one has to solve the strongly 
coupled theory in order to show that it is the same as some dual 
weakly coupled theory. In supersymmetric cases there are several quantities
that one can calculate which do not depend on the coupling. They can
be calculated at weak coupling, extrapolated to strong coupling, and then
compared with the corresponding result in the dual theory. 
For this reason dualities have been checked mostly in supersymmetric
theories. Examples of quantities that are protected by supersymmetry and
can be calculated are: 1) Low energy effective action, 2) The number
and masses of various ``protected'' special states, these are typically
some charged particles. These states could be elementary on one
theory and solitons on the dual theory. 
Books on string theory include \cite{polchinski,gsw}

\section{Black hole entropy}

\begin{figure}[htb]
\begin{center}
\epsfig{file=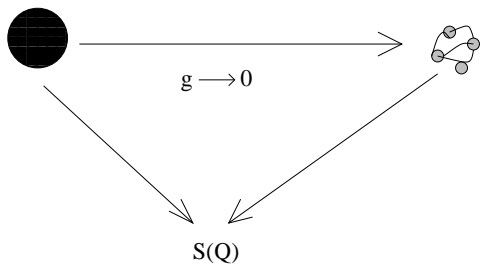}
\caption{Starting from an extremal charged black hole we change the 
value of the coupling and the black hole becomes a weakly coupled gas of
D-branes
and strings. In supersymmetric theories these are BPS states and their
number
does not depend on the coupling, so we can calculate the black hole entropy 
by counting the number of states in the gas of D-branes and strings 
\cite{Strominger:1996sh}. }
\label{fig:entropy}
\end{center}
\end{figure}

Black holes are one of the most intriguing objects that 
general relativity predicts.
In classical general relativity black holes have a horizon, 
which is a surface in
spacetime such that if somebody crosses it he/she  cannot come back out.
This surface, however,  does not 
look special at all for
the observer who is falling in.  
Black holes became even more intriguing when Hawking showed 
that they 
emit thermal radiation due to quantum effects. So a black 
hole is more like a black body, 
which will emit thermal radiation. For a black hole in four
 dimensions the 
temperature is inversely proportional to its radius,
 $T\sim 1/r\sim 1/(G_N M) \sim 10^{-4} K ( M_{sun}/M)$. 
We see that for black holes that are produced for normal
 astrophysical processes,
whose mass is always bigger than the mass of the sun, 
this temperature is too small
to be detected, since it much lower that the 
temperature of the cosmic microwave background radiation. 
The fact that they are thermal objects raises very 
interesting and very important theoretical puzzles,
solving these puzzles is one of the challenges of a theory of 
quantum gravity.
We are used to the fact that when we encounter a thermal object 
we can explain its
temperature as arising from the motion of the internal constituents. 
So the question becomes,
what are the internal constituents of the black hole that explain 
its temperature? This question is
often phased in terms of explaining the microscopic origin of the 
entropy. The entropy can be
defined through the first law of thermodynamics as $ dM = T dS$. 
The entropy comes of
to be $ S = A_H/(4 G_N)$. In other words, the entropy is 
proportional to the area of the horizon in Planck units. 
  Any theory of quantum gravity, such as 
string theory,
 should explain this entropy.  In string theory it is hard to 
calculate this entropy directly since
strings describe small fluctuations around flat space while a 
black hole represents a large 
deviation from Minkowski space.  Recently, when the dynamics of 
D-branes was understood, 
it became possible to calculate this entropy for some special cases 
\cite{Strominger:1996sh}.
  Consider a compactification 
of string theory down to four dimensions that preserves two 
supersymmetries. In such a theory 
we could consider charged black holes. In general charged 
black holes  should satisfy 
a constraint on the mass that looks like $M \geq Q$ in order 
to avoid naked singularities, i.e. singularities
which are not covered by a horizon.
In these theories with two supersymmetries this constraint 
coincides with the so called BPS bound. This is
a bound coming from the supersymmetry algebra. The charge $Q$ 
appears in the right hand side of the 
supersymmetry algebra and the BPS bound comes from demanding 
unitarity of the representations. 
Furthermore, the states with  $M=Q$ lie in a smaller representation 
of the supersymmetry algebra
and the number of states in such representations does not depend 
on the coupling or other continuous 
parameters in the theory\footnote{Most precisely  the number of 
states that cannot be combined 
into  larger representations, such as the ones with $M>Q$,  
remains invariant.}.
Black holes with $M=Q$ are also special from the point of
 view of the gravity theory, they are called 
extremal black holes and for them the Hawking temperature 
vanishes. 
In these supersymmetric theories it is possible to change 
parameters so that the black hole configuration
becomes a weakly coupled system of D-branes and strings whose 
entropy one can calculate fairly easily, see figure \ref{fig:entropy}. 
The answer, of course, comes out to be the same as the area of 
the corresponding black hole solution. 
Since the number of BPS states does not change when we do this 
transformation this provides a 
derivation of black hole entropy for these special black holes 
in these supergravity theories. 
The entropy of more general black holes, including near extremal 
black holes ($ M > Q$ but $M-Q \ll Q$),
can be computed using the AdS/CFT correspondence as it will be 
explained below.  The entropy of
general black holes in completely general  string backgrounds 
cannot be calculated with the present techniques. 

\section{Conformal field theories and Anti-de-Sitter spacetimes}

\begin{figure}[htb]
\begin{center}
\epsfig{file=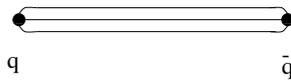}
\caption{Fluxes  of the color electic field forms a narrow bundle leading to a
linear
potential and confinement.}
\label{fig:flux}
\end{center}
\end{figure}

\begin{figure}[htb]
\begin{center}
\epsfig{file=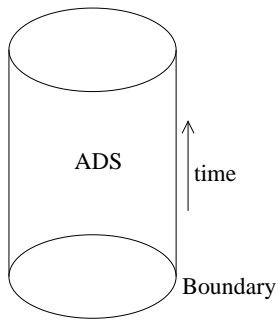}
\caption{Penrose diagram of Anti-de-sitter which shows its causal
structure. It is a solid cylinder. Time runs in the vertical direction. The
boundary of AdS
is the boundary of the cylinder. In the case of $AdS_5$ the boundary
is $R\times S^3$. }
\label{fig:penrose}
\end{center}
\end{figure}

Although string theory was described above as a theory of quantum 
gravity, it originated as an attempt to 
describe hadrons. The string description  explained some features 
of the hadron spectrum 
such as Regge trajectories, etc.. We now know that hadrons are 
described by QCD, but it is still quite hard
to do computations at low energies due to strong coupling problems.
 In fact we expect confinement. 
Confinement is thought  to arise from the fact that the color 
electric field lines form narrow bundles in going
from a quark to an anti-quark, see figure \ref{fig:flux}. 
These fluxes look at low energies 
like strings and one might expect that
at low energies a description in terms of string might be valid.  
It was shown by 't Hooft \cite{'tHooft:1974jz} that
the proper way to understand these strings is to take the limit 
of a large number of colors $N\to \infty$, 
keeping $g^2_{YM} N$ fixed. In this limit only planar Feynman 
diagrams survive. These planar Feynman diagrams seem to be  
giving a discretization of a string worldsheet. However it was 
not clear what kind of
string theory that would be. 
For the case of Yang Mills theory with ${ \cal N} =4 $ supersymmetries 
and a large number of colors $N$ 
it has been conjectured that these gauge strings are the same as 
the fundamental strings described above
but moving in a particular curved spacetime:  the product of 
five-dimensional Anti-de-Sitter space and a five-sphere
\cite{Maldacena:1998re}.
Five dimensional AdS has a  boundary which is four dimensional.
The field theory is defined on this four dimensional boundary.  
In figure \ref{fig:penrose} we can see the Penrose diagram of AdS.

\begin{figure}[htb]
\begin{center}
\epsfig{file=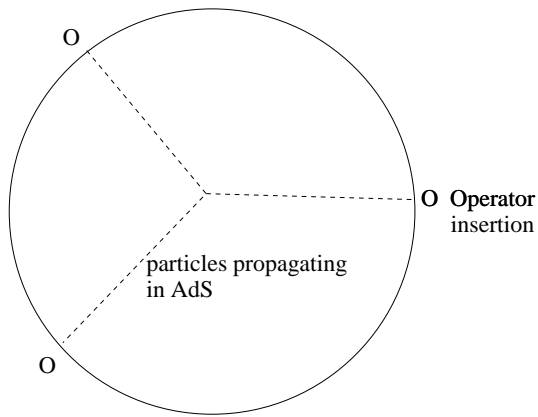}
\caption{The correlation function of operators in the boundary theory can
be 
calculated as the probability amplitude that the particles associated to
the operators travel between the  points at the boundary where the
operators are inserted.}
\label{fig:correlations}
\end{center}
\end{figure}

 The radius of curvature of this spacetime is 
proportional to $(g^2_{YM}N)^{1/4}$ and the string coupling is 
$g_s = g^2_{YM}$ which goes like $1/N$ for fixed $g^2_{YM} N$ 
as expected from 't Hooft's general argument. 
There have been a large number of checks for this correspondence.
 Many checks are possible due to the 
large number of supersymmetries. 
The simplest check is the observation that both theories have 
the same symmetries. ${\cal N}=4$ supersymmetric
Yang Mills theory is scale invariant, the coupling does not run, 
it is independent of the energy. 
Non supersymmetric Yang Mills is classically conformal invariant 
but quantum corrections introduce 
a length scale through the running of the coupling.  In conformal 
theories the symmetry group
includes translations, rotations, scale transformations and the 
so called special conformal transformations. 
All these form the group mathematically known as $SO(2,4)$. This 
group is the group of isometries of 
AdS.  Similarly the Yang Mills theory has an $SO(6)$ global 
symmetry group, which is the same 
as the group of rotations of $S^5$. In fact when we consider 
string theory on $AdS-5 \times S^5$ we also 
have the same supersymmetries as the gauge theory.  
Other checks include the comparison of the spectrum of BPS 
particles, renormalization group flows that partially break 
supersymmetry, etc.   
Many extensions of this correspondence were suggested, for 
other theories with less or no supersymmetry and for theories 
that are not conformal invariant. For a detailed account of the 
checks and a more extensive list of references on this subject
see \cite{Aharony:1999ti}. 
A puzzling aspect is that the bulk theory contains gravity 
while the field theory does not contain gravity. 
Gravity in the bulk is related to the 
stress tensor of the boundary theory in the following way. 
Correlation functions for the stress tensor  in the field 
theory are equated with  the amplitude for
propagation of gravitons between prescribed points at the 
boundary. 
Correlation functions of operators in the field theory 
can be calculated, using the correspondence,
as the amplitudes that particles propagate through the bulk between 
prescribed points at the boundary \cite{Gubser:1998bc,Witten:1998qj},
see figure \ref{fig:correlations}.
Each operator corresponds to a particle, or more precisely a string 
mode, propagating in $AdS$. The mass of 
the particle in AdS is related to the conformal dimension $\Delta$  
of the field through
$$  \Delta  = 2 + \sqrt{ 4 + ( m R_{AdS})^2 } $$
\begin{figure}[htb]
\begin{center}
\epsfig{file=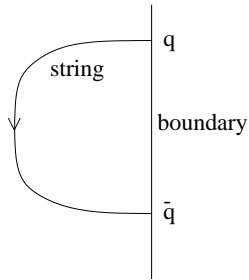}
\caption{A quark anti-quark configuration. The flux of color  field
joins the quark and anti-quark and it is described by a string that moves
in the higher dimensional geometry which is dual to the field theory. 
The potential can be calculated as the proper energy of this string in the
curved
geometry. }
\label{fig:wilson}
\end{center}
\end{figure}
Similarly we can calculate the the quark antiquark potential by 
considering a string that goes between 
two points on the boundary see figure \ref{fig:wilson}.  
In this picture we can see that the 
strings that move in the ten dimensional
space are precisely the strings representing fluxes of color  
fields. We can thus say that strings made of
gluons look very much like ordinary fundamental strings when 
$g^2_{YM} N$ is large. 
It is a general feature of dualities that some excitations that 
look fundamental in one picture are
collective excitations in the dual theory. 
If we compute the quark-antiquark potential by computing the 
energy of the string in this curved background
 we find that it goes like $V(L) \sim -1/L $ where $L$ is the 
separation in the four dimensional theory. 
This potential is not confining and it should not be, since the 
field theory is conformal. 
It is possible to deform the field theory in such a way that one 
destroys conformal invariance and supersymmetry at low energies. 
The resulting theory is expected to be confining. Indeed one can 
find the 
corresponding supergravity solution and find that the geometry is 
deformed and now the quark-antiquark 
potential is indeed confining  and the theory has a mass gap 
\cite{Witten:1998zw}.
 Even though the theory is confining it is not pure Yang-Mills, it is 
a strongly coupled version of it. In order to find the large $N$ 
limit of pure Yang-Mills one needs to
consider strings propagating in a curved spacetime whose curvature 
is of the order of the string scale. 
In this situation the gravity approximation would not be good enough. 
 Treating strings is these 
small spaces is a challenging problem, which is being explored. 
What we have seen so far are applications of the correspondence 
to the study of large $N$, strongly coupled field theories.  The 
correspondence can also be used to learn about gravity. It is 
harder to use it in this direction since the field theory is strongly 
coupled and therefore hard to solve. There are, however, some 
general statements that one could make. 
One of the most mysterious objects in a gravity theory is a black hole.
 One can consider a black hole in $AdS$. This black hole is, 
in principle, described by some thermal state in the boundary theory. 
If we have a big black hole, of the order of the size of the $AdS$ radius, 
its entropy will  be inversely proportional to the Newton constant 
which is of order $1/N^2$. So the entropy will be of order $N^2$, 
which agrees with the number that we would naively expect in the 
field theory since we have $N^2$ gluons. 
In the case of $AdS_3$ it is possible to precisely calculate the 
entropy in the field theory and the result agrees with gravity
 precisely, see the review  \cite{Aharony:1999ti}. 
An aspect of gravity that is manifest in via this  correspondence is 
holography \cite{'tHooft:1993gx,Susskind:1995vu}. 
Holography says that in a quantum theory 
of gravity we should be able to describe physics in some region 
of space by a theory with at most one degree of freedom per unit
Planck area. Notice that the number of degrees of freedom
would then increase  with the area and not with the volume as we are normally 
used to. Of course, for all physical systems that we normally  encounter
 the number of degrees of freedom is much smaller than the 
area, since the Planck length is so small. 
It is called ``holography'' because it would be analogous to a hologram
which can store a three dimensional image in a two dimensional surface. 
In this case we represent the physics of the five dimensional 
Anti-de-Sitter spacetime with a theory that lives on its boundary. It is
a concrete example of holography. Understanding it better might lead to 
more insights about quantum gravity.

\section{Conclusions}

We have learned that the laws of physics as we know them are 
not consistent, since we do not treat
gravity using quantum mechanics. Formulating the correct theory 
of quantum gravity would enable us
to understand the big bang and most probably the parameters of 
the standard model. String theory is
enables us to study   a whole 
class of phenomena that we expect  in a theory of quantum gravity. 
The theory is not yet developed to the point of making definite 
experimental predictions,  but is 
understood well  enough to explain some quantum gravity phenomena like 
black hole entropy and evaporation, topology change, etc. 
Supersymmetric phases of string theory are fairly well explored. 
One of the main challenges is to find
a supersymmetry breaking mechanism that does not generate a large
 cosmological constant. 
Many connections were found between different string theories and 
between string theory and field theory. 
It has been shown that string theory reduces, in certain circumstances, 
to ordinary four dimensional field theories and vice-versa.
We hope that in the near 
future we will  understand the theory better 
so as to make contact with experiment.


\bigskip

I am grateful 
to the organizers and participants for an interesting conference. 
This work was supported in part by DOE grant DE-FG02-91ER40654 and the 
Sloan and Packard foundations.

\end{document}